\documentstyle[preprint,aps,epsf]{revtex}

\tighten
\begin{document}
\draft
\preprint{TPI-MINN-99/17, $\;\;$ UMN-TH-1755-99}

\newcommand{\nc}{\newcommand}
\nc{\al}{\alpha}
\nc{\ga}{\gamma}
\nc{\de}{\delta}
\nc{\ep}{\epsilon}
\nc{\ze}{\zeta}
\nc{\et}{\eta}
\renewcommand{\th}{\theta}
\nc{\Th}{\Theta}
\nc{\ka}{\kappa}
\nc{\la}{\lambda}
\nc{\rh}{\rho}
\nc{\si}{\sigma}
\nc{\ta}{\tau}
\nc{\up}{\upsilon}
\nc{\ph}{\phi}
\nc{\ch}{\chi}
\nc{\ps}{\psi}
\nc{\om}{\omega}
\nc{\Ga}{\Gamma}
\nc{\De}{\Delta}
\nc{\La}{\Lambda}
\nc{\Si}{\Sigma}
\nc{\Up}{\Upsilon}
\nc{\Ph}{\Phi}
\nc{\Ps}{\Psi}
\nc{\Om}{\Omega}
\nc{\ptl}{\partial}
\nc{\del}{\nabla}
\nc{\be}{\begin{eqnarray}}
\nc{\ee}{\end{eqnarray}}
\nc{\slc}{$SL(2,${\bf C}$)\,$}
\nc{\htt}{\hat{\ta}}
\nc{\qh}{\hat{q}}

\title{On Modular Invariance and 3D Gravitational Instantons}
\author{Thorsten Brotz$^{1}$, Miguel E. Ortiz$^{1}$,  
        and Adam Ritz$^{2}$\footnote{aritz@mnehpw.hep.umn.edu}}
\address{$^{1}$ Theoretical Physics Group, Blackett Laboratory \\
         Imperial College, Prince Consort Rd., London, SW7 2BZ, U.K.\\
         $^{2}$ Theoretical Physics Institute, University of Minnesota \\
         116 Church St., Minneapolis, MN 55455, USA}
\date{\today}

\maketitle

\begin{abstract}
We study the modular transformation properties of Euclidean solutions
of 3D gravity whose asymptotic geometry has the topology of a torus.
These solutions represent saddle points of the grand canonical partition 
function with an important example being the BTZ black hole, and their
properties under modular transformations are inherited from the 
boundary conformal field theory encoding the
asymptotic dynamics. Within the Chern Simons formulation, classical 
solutions are characterised by specific holonomies describing the 
wrapping of the gauge field around cycles of the torus. We find that
provided these holonomies transform in an appropriate manner,  
there exists an associated modular invariant grand canonical 
partition function and that the spectrum of saddle points naturally includes a 
thermal bath in $AdS_3$ as discussed by Maldacena and Strominger. 
Indeed, certain modular transformations can naturally be described 
within classical bulk dynamics as mapping between different foliations 
with a ``time'' coordinate along different cycles of the asymptotic torus. 
\end{abstract}

\vfill\eject

\section{Introduction}
 
One of the regimes in which the correspondence \cite{malda,gkp,witten}
between string theory on anti-de Sitter space ($AdS$) and 
conformal field theory is under some
measure of control is in the case of $AdS_3$. In this example, there
are no propagating gravitational modes in the bulk and it has been known for
some time \cite{bh} that the asymptotic dynamics of the gravitational
field, with appropriate boundary conditions, induces a local
conformal symmetry with central charge,
\be
 c_{cl} & = & \frac{3l}{2G}, \label{ccharge}
\ee
in which $G$ is Newton's constant and $-2/l^2$ is the cosmological
constant. The particular asymptotic fall-off required as a boundary condition
for the metric, for preservation of the asymptotic isometry, 
is known \cite{bh} to be equivalent to the conformal
treatment of infinity of Penrose, and which was discussed in the
context of the $AdS/CFT$ correspondence by Witten \cite{witten}. 

As a consequence of the trivial bulk dynamics, it is well known that
under these particular boundary conditions, the asymptotic dynamics of the
gravitational field effectively reduces to classical Liouville 
theory \cite{chd} (see also \cite{weyl}). Then, as discussed by
Martinec \cite{martinec2}, the $AdS/CFT$ correspondence
for the metric--stress tensor sector may be rephrased in this example
as the semiclassical relation 
$T_{zz}^{Liouville}=\left<T_{zz}^{CFT}\right>$,
or an appropriate generalisation to include correlators. 
Liouville theory then describes the effective action for the 
Weyl anomaly of the CFT arising from its coupling to the background
geometry, encoded in the conformal Ward identities. In the approach of
\cite{martinec2}, the level density of Liouville theory, 
equivalent to that of a free boson, i.e. $c_{eff}=1$, is only a thermodynamic
description of the microstates underlying 
the CFT associated with ({\ref{ccharge}), and which would presumably also 
describe the microscopic origin of Bekenstein-Hawking entropy of 
the BTZ black hole following the argument of \cite{strominger}.

The suggestion that the asymptotic gravitational dynamics encodes
(or is induced by) the current sector of an underlying CFT presents
a new avenue to investigate inter-relations between classical bulk 
geometries and 2D CFT notions such as uniformization theory and
modular invariance. The first of these was investigated in \cite{martinec2}.
Recall that classical 3D Euclidean geometries may be realised as
a quotient of Euclidean $AdS_3$ by an element of \slc, and that these 
matrices are equivalent up to conjugation to the holonomies of the gauge
field describing the same classical solution within the
Chern Simons formulation of 3D gravity \cite{ct}. Since the asymptotic
boundary is necessarily a Riemann surface, it was shown in 
\cite{martinec2} that there is a correspondence between the conjugacy
classes of holonomies and particular uniformizing coordinates associated with
classical Liouville fields living on the boundary. 

The advantage of working with $AdS_3$ is that one may go beyond the 
semi-classical formulation of the $AdS/CFT$ correspondence, working
with small curvatures $l\gg 1$, and consider quantum corrections in the
bulk, and thus subleading $O(1/c_{cl})$ corrections in the boundary CFT.
The second point noted above, the manifestation of modular 
invariance in terms of classical bulk geometries, which will be the
subject addressed 
in the present work, requires an extension of this kind. 
As motivation, we note that in Euclidean
signature, the compactified boundary of $AdS_3$ may be interpreted
as a Riemann sphere. In practice, one does not
need to assume this theory actually lives at the boundary, but 
keeping this analogy one may recall that the presence of a CFT
on the plane generally implies the existence of a modular invariant
theory on the torus \cite{mod}. Indeed, an asymptotic torus geometry
is known to be the topology associated with the Euclidean BTZ
black hole \cite{btz}. The question then arises as to whether
the BTZ black hole has an associated modular invariant partition function,
or indeed how modular transformations would manifest themselves
in the bulk theory.

To explore this suggestion in more detail, we
concentrate for the moment on the BTZ black hole \cite{btz}.
We may write the action in terms of standard thermodynamic quantities, 
the average energy $\left<E\right>$, spin $\left<J\right>$ , and 
entropy $S$, as $I=\beta\left<E\right>+\Om\left<J\right>-S$. 
The classical black hole partition function, $e^{-I}$, then takes
the form
\be
 e^{-I_{BH}} & = & \exp\left(2i\pi\frac{c_{cl}}{24}\left(\frac{1}{\ta}-
     \frac{1}{\overline{\ta}}\right)\right),
\ee 
where $\left<E\right>=M$ is the black hole mass, $\beta$ is the inverse
temperature, $\Om$ is the angular velocity, and $\ta$ is a modular
parameter (complex structure), given by \cite{bbo}
\be
  \ta & = & \frac{\beta}{2\pi}\left(\Om+\frac{i}{l}\right). \label{tau}
\ee
In a recent analysis, Maldacena and Strominger \cite{ms}
noted that another saddle point, which one may interpret as a thermal
bath in $AdS_3$, may be represented by the action $I=\beta\left<E\right>$,
and has a classical partition function of the form
\be
 e^{-I_{TAdS}} & = & \exp\left(-2i\pi\frac{c_{cl}}{24}(\ta-\overline{\ta})
    \right),
\ee
where $\left<E\right>$ is now interpreted as the
$AdS_3$ mass, which importantly is negative. 
This re-interpretation naturally corresponds
to the rewriting of the full ensemble above via use of the
3D Smarr formula $-(\beta M+\Om J)+S=(\beta M+\Om J)$. It was
observed in \cite{ms} that these two solutions are related
by the modular transformation $\ta\rightarrow -1/\ta$, and that this
transformation corresponds to an analogous mapping in the 
element of $SL(2,${\bf C}$)$ by which they were constructed,
leading to a conjecture that one should have a full $SL(2,${\bf Z}$)$ 
spectrum of such solutions. This is plausible since both solutions
above have the geometry of a solid torus, while the radial coordinate
actually decouples from the dynamics. Thus the effective theory
is defined on a torus and, if conformal as understood above, one expects
a modular invariant partition function to exist. Indeed it was noted
in \cite{ms}, that the dual conformal field theory was defined
on the complex plane with specific identifications analogous to the
identifications used in constructing the bulk solutions. 

Since this modular invariant structure is associated with transformations
of the quotient $SL(2,${\bf C}$)$ matrix, one expects that these features 
should also be apparent in the Chern-Simons formalism as particular 
transformations
on the holonomies of classical solutions. In this paper we shall
show that this is indeed the case, and that a natural
way to picture the effect of modular transformations such as
$\ta\rightarrow-1/\ta$ at the classical level is as a mapping to new
canonical 2+1 decompositions of the Chern-Simons action, where the
``time'' direction is associated with different cycles of the asymptotic
torus. On the other hand, at the level of the CFT,
the holonomies of the classical gauge field naturally enter via
an appropriate choice of basis for the asymptotic symmetry algebra.
The modular invariance of the grand canonical partition function \cite{bbo}
is conveniently made manifest by working with the chiral WZW models 
associated with the gauge fixed Chern Simons action. This choice of
basis then enters through the characters of the Kac-Moody algebra
describing the WZW partition function. 

The layout of the paper is as follows. In Section 2 we review the structure
of asymptotically toroidal solutions in 3D gravity, and construct a generic
2+1 decomposition in first order form. In Section 3 we discuss the
prominent role played by the defining holonomies in determining 
the modular invariance properties of the grand canonical ensemble.
We finish in Section 4 with some additional remarks on the semi-classical 
limit. 

\section{General Canonical Decompositions}

Since classical solutions of 3D gravity have
constant curvature, they can be represented in terms of a 
quotient of the hyperbolic space (Euclidean $AdS_3$)
\be
 l^{-2}ds^2 & = & e^{2\rh}dwd\overline{w} + d\rh^2,
\ee
by an element $H$ of \slc, where $w$ is a complex coordinate on the plane,
the radial coordinate is $\exp(\rh)$, and $l$ denotes the scale of 
the geometry. In particular, we shall be concerned here with 
asymptotically toroidal geometries, so that $H=H(\ta)$ is diagonal,
and in the case of the BTZ black hole, given by
\be
 H & = & \left(\begin{array}{cc}
                    e^{-i\pi/\ta} & 0 \\
                    0 & e^{i\pi/\ta}
               \end{array}\right).  \label{identmat}
\ee
This generates the identifications,
\be
 && \rh \sim \rh +i\pi\left(\frac{1}{\ta}+\frac{1}{\overline{\ta}}\right),
   \;\;\;\;\;\; w \sim \exp\left(-2i\pi\frac{1}{\ta}\right)w.
\ee

In this representation, the identification of the black hole
boils down to the relation between $\ta$ and the physical variables
$\beta$ and $\Om$. In particular, were one to invert the identification
in (\ref{tau}), and identify these physical parameters with $-1/\ta$, then
the classical solution we have just described via identification
of the hyperbolic space would correspond to a thermal bath in
$AdS_3$ \cite{ms} as discussed in the introduction. Of course, it is
equivalent to retain the identification of $\ta$ in (\ref{tau}) and instead
to redefine the identification matrix (\ref{identmat}). This ambiguity
in interpretation
re-enters when we consider the appropriate choice of boundary conditions
below.

The matrix $H$ has a very natural interpretation in the Chern-Simons
formulation. To see this, we now recall that the
Einstein--Hilbert action for Riemannian geometries, discarding
invertibility for the dreibein (see \cite{wit91}), may be 
rewritten as the difference of two Chern-Simons actions, with
gauge group $SL(2,${\bf C}$)$ \cite{cs}, 
\be
 I_{EH} & = & \frac{1}{16\pi G}\int_M\sqrt{g}\left(R+\frac{2}{l^2}\right)
     \;\; = \;\; i\left(I[A]-I[\overline{A}]\right) + B,
\ee
where $B$ is a boundary term. As noted in \cite{maxrev}, the first-order
form has a sign ambiguity associated with the relation $\sqrt{g}=\pm e$,
where $e$ is the determinant of the dreibein. The choice
fixes the relative orientation of the orthonormal and metric frames,
and we shall find it convenient to choose the (unconventional)
negative sign (and $e<0$), so that the Chern-Simons
level $k$, to be introduced shortly, will be positive. The gauge fields 
are then defined in terms of the spin connection and dreibein
as $A^a = \om^a+ie^a/l$, and $\overline{A}^a=\om^a-ie^a/l$. 
The gauge field configuration corresponding to 
the Euclidean black hole is given by 
\be
 A & = & -\frac{1}{2}\left(\begin{array}{cc}
                   dr & ie^{r}dz/\ta \\
                   ie^{-r}dz/\ta & -dr
               \end{array}\right),  \label{gauge}
\ee
where $r$ is a proper radial coordinate, and $z=\varphi+\ta x^0$ in terms
of the conventional real, periodic, Euclidean black hole coordinates. 
We note that the coordinate $z$
naturally lives on a cylinder and is related to the plane coordinate
$w$ by an exponential mapping, the explicit form of which we shall not 
require. The crucial feature we wish to emphasize, is that the 
global structure of the gauge field is encoded in the identification
matrix $H$ (\ref{identmat}). This follows from the holonomies of (\ref{gauge})
which are given by
\be
 P\exp\left(\int_{\ga}A\right) & = & M H M^{-1},
\ee
where $M$ is an \slc matrix. Consequently, we should expect that the
modular structure of classical solutions will involve the gauge field
holonomies in a direct manner.

As we consider spacetimes which have the geometry of a solid torus,
there are two natural choices for the foliating coordinate to use
in pursuing a 2+1 canonical decomposition, namely, the contractible and
non-contractible cycles. For generality we shall denote the
choice of foliation coordinate as $u$, and the coordinate along the
other cycle as $v$.
Now, in formulating a path integral representation for the  
grand canonical partition function, one needs to specify boundary conditions
which ensure that variations of the action are well defined, while
also defining the ensemble by fixing certain physical quantities.
This issue has recently been considered in some detail in \cite{bm}.
The main point to emphasize is that for the ensemble of 
interest, the covariant form of the Chern Simons action requires no 
boundary terms at all!

In order to specify the boundary conditions, we now concretely
define $\ta$ in terms of physical quantities 
via (\ref{tau}), and reflect different identifications
of the hyperbolic space, via the \slc  mapping 
$\htt=(a\ta+b)/(c\ta+d)$. Then we have $z=v+\htt u$
and the boundary conditions appropriate to this problem are chiral,
\be
 && A^a_{\overline z} = 0, \;\;\;\;\;\;\;
  \overline{A}^a_z = 0. \label{bcs}
\ee
An important consequence of these relations, as noted in
\cite{carlip97} and fully revealed in \cite{maxrev}, is that 
this condition links the real and imaginary
parts of $A$ at the boundary, where the nonzero components are
given by
\be
 && A^a_z = 2 \om^a_z, \;\;\;\;\;\;\;
  \overline{A}^a_{\overline z} = 2\om^a_{\overline z}.
\ee
Consequently, the \slc gauge field reduces at the boundary to the two real
$SU(2)$ currents above.

Since the covariant Chern-Simons action 
requires no boundary terms, the general 2+1 decomposition 
with $A=A_udu+A_idx^i$ then takes the form,
\be
  I[A,\ta] & = & \frac{k}{4\pi}\int du \int_{\Si}\ep^{ij}Tr\left(-A_i\ptl_uA_j+
        A_uF_{ij}\right)\pm\frac{k\htt}{4\pi}\int_{T^2_{\infty}}
                TrA_{v}^2, \label{2+1gen}
\ee
where $k=l/(4G)=c_{cl}/6$ so that with $e<0$ both $k$ and $c_{cl}$ are
positive. In fact, we should mention that there is a 
subtlety\footnote{We thank M. Ba\~nados for remarks on this point.} if 
one chooses $u=x^0$, in that the foliation becomes degenerate at $r=0$.
This can be treated by removing a small disk $r<\ep$ on which the covariant
action is well defined. However, we shall find here that an additional
source will necessarily arise at $r=0$, and resolve the degeneracy.
Thus, given this caveat, we shall work generically with (\ref{2+1gen}).

The sign of the boundary term depends on the choice of $u$, as a 
consequence of the antisymmetry of the boundary 2-form. In this
form, one takes the positive sign when $u$ denotes the non-contractible
cycle.

This representation is simply convenient for compactly expressing
results associated with different solutions and different foliating
coordinates. In order to have some specific examples we quote below
the boundary conditions for the on-shell BTZ black hole and also the
thermal $AdS$ bath ($TAdS$),
\be
 &&\mbox{BTZ} = \left\{\begin{array}{cc}
                      u=x^0, & \htt=\ta \\
                      u=\varphi, & \htt=\frac{1}{\ta}
                         \end{array}\right. \;\;\;\;\;\;\;\;\;\;\;\;\; 
 \mbox{TAdS} = \left\{\begin{array}{cc}
                      u=x^0, & \htt=-\frac{1}{\ta} \\
                      u=\varphi, & \htt=-\ta
                         \end{array}\right.. \label{tadsbc}
\ee
Henceforth we shall focus our attention on these two Euclidean saddle
points as they have a natural interpretation. For a specific choice
of $u$, the transformation between boundary conditions for each saddle point
is just $S:\ta\rightarrow-1/\ta$ as one would expect from the discussion
of section 1. Note also that if we switch between cycles in defining
$u$, we also switch between the two saddle points up to a sign.
This sign change is accounted for by an analogous change in the
boundary term, and thus the form of the action is equivalent.
 This is not surprising since, as discussed earlier,
the geometry of these solutions as described by the identification
matrix $H$ is equivalent up to a modular transformation of $\ta$.
However, the semiclassical regime in each case
is quite different, with large black holes corresponding to the
limit $|\ta|\ll1$, while a low-temperature thermal bath corresponds
to $|\ta|>>1$. As a consequence it was conjectured in \cite{ms} that
since these solutions are Euclidean saddle points in the partition function
there is a phase transition at some temperature separating phases
dominated by each saddle point. We shall show in the next section that
there is a natural grand canonical partition function which is indeed
modular invariant and thus contains all $SL(2,${\bf Z}$)$ related
saddle points.

\section{The Partition Function and Modular Invariance}

Working within the class of asymptotically toroidal geometries we 
have the crucial simplification that the \slc gauge field reduces
at the boundary to two real affine $SU(2)$ currents at 
level $k$, whose components we denote again by
$\{A^i\}$ and $\{\overline{A}^i\}$, with $i=1,2,3$
and the $SU(2)$ conventions are
$[T_a,T_b]=i\ep_{abc}T^c$ with $Tr(T_aT_b)=\de_{ab}/2$.

For each of these fields, variation of $A_u$ implies the constraint
$F_{ij}=0$, the most general solution of which, for a
manifold with nontrivial cycles such as the solid torus,
is given by \cite{emss}
\be
 A & = & g^{-1}dg+g^{-1}\th(u,v)g \;\; \equiv \;\; 
   (gg_0)^{-1}d(gg_0).\label{decomp}
\ee
In this expression $g$ is a single-valued group element, 
and $\th$ is a Lie algebra-valued one-form which need depend only
on the coordinate parametrising the contractible cycle.
With the choice of gauge used
in \cite{bbo} and consistent with the boundary conditions of \cite{bh}, 
the $\rh$--dependence may be factored out of the
group element and the exterior derivative above reduces to a single
$\ptl_v$ component. Note, however, that this is a weaker constraint
than that required to reduce the asymptotic dynamics to Liouville theory
\cite{chd}. Since $\pi_1(T^2)=${\bf Z}$\oplus${\bf Z}
is commutative, the components of $\th$ may be rotated into the
Cartan subalgebra of the group ($T_3$). Under ``large'' gauge transformations
(modular transformations) $\th$ can change since it encodes the
holonomy of the gauge field, and thus measures the wrapping around the
non-contractible cycle of the torus.
This zero mode will play a vital role in determining the 
value of the saddle point, and indeed the modular invariance properties, 
so its now convenient to fix $u$ equal to the contractible cycle, so that
$\th$ is present in the ``spatial'' slices. We shall comment on the
alternative choice in Section~4.

In the conventional, operatorial, quantisation of Chern-Simons theory
the zero-mode factor $g_0$ (or $\th(u)$) may be parametrised 
as an appropriately normalised Wilson line. As we noted above, the 
zero-mode sector of the gauge field may be rotated into the Cartan subalgebra,
and thus this Wilson line may be represented as
\be
 g_0 & = & P\exp\left(const\times\int_{\ga(u,v)}\om(u,v)a
                \cdot T^3\right),
\ee
where $a$ is a constant, and $\om$ is the abelian differential of the
torus. For concreteness, if one chooses the holomorphic coordinates
$(z,\overline{z})$, then the Lie algebra element may be written as \cite{lab},
\be
 g_0^{-1}dg_0 & = & \frac{ia}{2\rm{Im}\ta}\overline{\om}T^3
                   - \frac{i\overline{a}}{2\rm{Im}\ta}\om T^3. \label{zero}
\ee
The holomorphic abelian differential is normalised so 
that $\int_{\al}\om=2\pi$ and
$\int_{\beta}\om=2\pi\ta$ where $\al$ and $\beta$ are the contractible and
non-contractible cycles respectively on the solid torus.

For example, the presence of this zero-mode in the black hole solution may be
observed by noting that the on-shell gauge field has a non-vanishing
$T^3$ component at $r=0$,
\be
 A^3|_{r=0} & = & \frac{i}{2\rm{Im}\ta}
    (dz-d\overline{z}), \label{bhzero}
\ee
which has precisely the form of (\ref{zero}) with $a=\overline{a}=1$. Note
that the sign of $a$ is according to convention and can be ignored.

In the partition function we are considering, this zero-mode characterises
the particular class of classical solutions we are interested in, and thus
should be fixed\footnote{We note that 
quantisation of the zero mode as a quantum mechanical system was initially 
discussed in \cite{emss,wit91}. This is of course important in theories
without boundary, and see \cite{cn} (and references therein) for a recent 
discussion of modular invariance in this context.}. 
Proceeding in this manner,
the calculation of the partition function reduces to the
construction of a specific Chern-Simons state, i.e. with prescribed
holonomies around the cycles of the torus \cite{carlip97}. 
Note that while $A^3_0$ can be
set to zero by a globally well-defined (large) gauge transformation,  
we need to distinguish such modular transformations as they change
the nature of the classical solution, as discussed in the introduction.

With the zero-mode held constant in this way, one finds after solving the constraint,  two copies of a classical $SU(2)$ chiral WZW (CWZW) model for the 
group element $gg_0$ at level $k$. Thus the full partition
function factorises as
\be
 Z_{SL(2,C)}(\ta,\overline{\ta}) & = & |Z_{SU(2)}(\ta)|^2.
\ee
As is well known, the CWZW Lagrangian has the non-canonical 
form $L=l_a(a)\dot{x}^a-H(x)$, in which the symplectic structure
is just the $SU(2)$ Kac-Moody algebra,
\be
 [T_n^a,T_m^b] & = & i\ep^{ab}_cT^c_{n+m}+n\frac{k}{2}\de^{ab}\de_{n+m,0},
\ee
where $T^a_n$ are the Fourier components of the gauge field,
\be
  && A_{v} = g^{-1}\ptl_{v}g =
    \frac{2}{k}\sum_{n=-\infty}^{\infty}
                       T_n^ae^{inv}.   \label{Tn}
\ee
The Hamiltonian has the form
$H(\htt)=k\htt\mbox{Tr}(g^{-1}\ptl_{v}g)^2/(4\pi)$, and thus the
partition function reduces to the expectation value of the above Wilson line,
parametrising the zero mode, in the CWZW model characterised by the
Hamiltonian $H(\htt)$,
\be
  Z_{SU(2)}(\ta) & = & \left<W_R\right>_{CWZW}, \;\;\;\;\;\;\;\;
   \mbox{where}\;\;\;
   W_R = \exp\left(\frac{k}{2}\int_{S^1_{r=0}}a\cdot T^3\right),
\ee
which depends only on the representation $R$ since $a$ is fixed. The
normalisation follows by reference to the definition (\ref{Tn}).
One may now observe that this Wilson line is precisely what was
interpreted as an additional boundary term in the analysis of
\cite{bbo}. We see here that it arises naturally in the parametrisation
of the zero-modes of the gauge field.

The calculation of the partition function is then formally equivalent
to the calculation of a Chern-Simons state with a prescribed
Wilson line around the non-contractible cycle of the solid torus. 
For large $k$ (i.e. small curvatures) the $AdS/CFT$ correspondence
relates the classical saddle point for 
$Z_{SL(2,C)}(\ta,\overline{\ta})\sim e^{-I_{inst}}$
to the boundary CFT partition function. However, 
we can now consider the generic structure for finite $k$, and
given the factorisation into $SU(2)$ currents at the boundary,
one expects the partition function to be given by \cite{hayashi}
\be
 Z_{SL(2,C)} & = & \sum_{2s=0}^k \overline{\ch_{2s,k}(a,-\htt)}
   \ch_{2s,k}(a,-\htt),
         \label{Zfull}
\ee
where $\ch_{2s,k}(a,\htt)$ are the characters for affine $SU(2)$, and $s$
labels the spin of the representation. The shift in the sign of
$\htt$ arises from our initial choice of the frame orientation, $\sqrt{g}=-e$.
Recall that as a consequence of (\ref{bcs}) we have $A^a_z=2i e^a_z/l$
at the boundary, and thus the in using (\ref{Tn}) we are working 
with a minimum weight condition on states, $T_n^a|0\rangle=0$ $(n>0)$, 
rather than the usual maximum weight condition, $T_{-n}^a|0\rangle=0$. 
The sign reversal for $\htt$ is simply a convenient means to re-orient 
the generators so that we may avoid this subtlety in 
working with $SU(2)$ characters. 

That (\ref{Zfull}) indeed arises in this case may be seen explicitly from the
structure of the Hamiltonian $H=H(\htt)$, and recalling that one may
realise the conformal symmetry of the model
through the Sugawara construction,
\be
 L_n & = & \frac{1}{k_q}\sum_{m=-\infty}^{\infty}:T_m^aT_{n-m}^a:.
\ee
The Virasoro generators $L_n$ are the Fourier 
coefficients of the energy-momentum tensor $T(v)$ (or appear via
a Laurent expansion $T(w)=\sum_nL_nw^{-n-1}$ in the coordinates
$(w,\overline{w})$). We have introduced $k_q=k+c_v$, the quantum shifted 
level, where $c_v=2$ for $SU(2)$, which arises in regularisation of the 
composite operator
$A^2$. Then the $SU(2)$ partition function may be expressed
\cite{bbo} as a sum over representations of 
\be
 \ch_{2s,k}(a,\htt) & = & \mbox{Tr}_s (W_s) \qh^{L_0-c_q/(24)}\;\; 
      = \;\; \mbox{Tr}_s e^{2\pi i a T^3_0}\qh^{L_0-c_q/(24)},
\ee
where $\qh=\exp(2\pi i \htt)$,
which is indeed recognisable as the definition of an affine $SU(2)$ 
character, in a particular basis associated with the Weyl chamber of affine
$SU(2)$. To see this, recall that (see e.g. \cite{kac,gw})
for an affine group, the root space is spanned in addition to the 
root space of the finite Lie group, by two additional generators
$\La_0$ and $\de$. This is because the maximal commuting set of 
generators now includes, along with the Cartan subalgebra
of $SU(2)$, the canonical central element $c$, and the number operator
$d=-L_0$. We denote the basis vectors associated with these generators
as $\La_0$ and $\de$. Then, since $SU(2)$ has just one simple root,
with a basis vector $\nu$, we can decompose any vector $V$ in root space
as,
\be
 V & = & 2\pi i (a\nu-\ta\La_0-b\de). \label{basis}
\ee
The notation we have used is consistent with our construction above, since
we recall that the character of a representation is given by
$\ch_R\equiv \sum_{\la}\rm{dim} R_{\la}e^{\la}$ where $\rm{dim} R_{\la}$ is the
multiplicity of the weight $\la$ in the representation $R$. Thus
we see that $a$ gives the coefficient of $T^3$, while $\ta$ (or
$\htt$ in this case) is the 
coefficient of the number operator, $d=-L_0$. The final 
vector $b\de$ does not appear
in the character above, but we shall comment on this shortly.

Expressions for affine $SU(2)$ characters are well known
(see e.g. \cite{gko,bbo}) as are their properties under
modular transformations \cite{kac}. The feature which will be of
relevance here is the role played by the zero mode of the gauge field. 

We consider the action of the generators of 
$PSL(2,${\bf Z}$)=SL(2,${\bf Z}$)/${\bf Z}$_2$ which we denote 
$S$ and $T$. Their action on $\ta$ is given by $S: \ta\rightarrow -1/\ta$,
and $T: \ta \rightarrow \ta+1$. $\ch_{2s,k}$ is known to
transform under $T$ via a phase determined by the conformal dimensions
of the primary fields of the model \cite{gw,lab}. Of most interest here
is the transformation under $S$. One finds, with the particular
action on the basis (\ref{basis}) described in \cite{kac}, that
$\ch_{2s,k}(\htt,a,b)$ transforms as follows:
\be
 \ch_{2s,k}\left(-\frac{1}{\htt},\frac{a}{\htt},0+\frac{a^2}{4\htt}\right) 
  & = & 
    \sqrt{\frac{2}{k+2}}\sum_{2x=0}^{k}\sin\left[\pi\frac{(2s+1)(2x+1)}{k+2}
        \right]\ch_{2x,k}(\htt,a,0).
\ee
The particular form of the modular transformations we have used
deserves some comment. One sees that with $a=b=0$, one recovers
the standard action on $\htt$. The transformation on $a$ may be understood
by recalling that it describes the holonomy of the gauge field
around the non-contractible cycle of the torus. This is unchanged under
$T$, but the cycles flip under $S$, and thus for the holonomy to remain
invariant, recalling the normalisation of the abelian differential, one
requires the transformation above. In the introduction we noted that
the matrices generating classical solutions via a quotient
of the hyperbolic space are expected to be equivalent, up to conjugation
by group elements, to the holonomies of the gauge field. Thus it
is not unexpected that the parameter $a$ must transform under
modular transformations. 
Finally, the basis element associated with the final coordinate 
$b$ is associated
with the lowest eigenvalue of $L_0$, and thus the above shift
reflects a shift of this kind. We shall return to this again shortly.

We can now conclude that the structure of the transformation 
properties under $T$ and $S$ ensures
that $\ch_{2s,k}$ form a unitary representation of the modular
group and as a consequence $Z_{SL(2,C)}$, since it is given 
by (\ref{Zfull}) as a diagonal sum over the characters, will
be modular invariant \cite{gw,lab,hayashi}. Consequently, the result
necessarily contains an $SL(2,${\bf Z}$)$
spectrum of saddle points associated with its invariance under the
choice of $\htt=(a\ta+b)/(c\ta+d)$, provided we now restrict
$a,b,c,d\in${\bf Z}.

However, we should point out that this result relied on using 
a specific definition of
the modular transformation associated with transforming the 
coordinate $b$ from zero to $a^2/(2\htt)$. This coordinate does
not naturally appear in the black hole solution for example,
and thus it is more natural to assume that
its fixed to zero, and not modified by modular transformations. If this
definition is pursued one finds that modular invariance does not
hold due to an additional factor of
\be
 && \exp\left(2i \pi \frac{k}{4}\frac{a^2}{\htt}\right),
           \label{shift}
\ee
which arises in the transformation law of $\ch_{2s,k}(\ta,a)$. This
is the factor that was previously cancelled by the transformation of $b$.
In the operatorial approach to Chern-Simons theories, one usually
recovers modular invariance by the addition of a 
prefactor $\exp(\pi k a^2/(4\rm{Im}\htt))$ which one may readily check
restores modular invariance for the sum over representations
of the combination,
\be
  \chi^M_{2s,k}(\htt,a) & = & \exp\left(\pi \frac{k a^2}{4\rm{Im}\htt}\right)
           \ch_{2s,k}(\htt,a).
     \label{modchi}
\ee 
However, one observes from the form of 
(\ref{shift}), that such a term would be cancelled if the character
were to be defined in terms of $L_0+ka^2/4$ with $a$ fixed,
rather than $L_0$. With the value of $a=1$ associated with 
the on-shell black hole, one sees that this is a shift 
$L_0\rightarrow L_0+c_{cl}/24$ (cf. (\ref{ccharge})).

We can understand this relation as a straightforward "improvement term"
for the Virasoro generators. To see this we recall
that given the Sugawara definition for the Virasoro 
generators above,
it is then a standard argument based on conformal invariance
that a modular invariant partition function may be defined by 
the diagonal sum of Virasoro characters,
\be
 Z_V & = & \sum_{R,\overline{R}}|\ch_V|^2, \;\;\;\;\;\;
       \mbox{where}\;\;\;\;\; \ch_V = {\rm Tr}_R \qh^{L_0-c/24},
      \label{vspec}
\ee
where $R$ represents a highest weight module of the algebra, 
Importantly,
the latter relation corresponds to a specific specialisation of the 
character, corresponding to the choice of a basis
with a component in the direction of $L_0$. Note that in general the
character can be defined by taking the trace of the exponential
of a given element of the maximal abelian subalgebra. 

For the CWZW model, one finds that 
$c=c_q=k$dim$(G)/(k+c_v)=3k/(k+2)$, and thus in the semi-classical limit
we have $c_q\sim 3$ for $k\gg 1$. The asymptotic conformal symmetry
algebra discussed in the introduction, and associated with
our choice of boundary conditions, has $c_{cl}=6k$ in terms of the
$SU(2)$ level in the semi-classical limit,
and thus does not directly correspond to the construction above.
Indeed this is why quantising the WZW model (or alternatively
Liouville theory) cannot directly explain the microstates associated with
the asymptotic conformal symmetry. 

The relevant shift in the definition of the Virasoro generators
so that they realise the asymptotic conformal symmetry
with central charge $\sim 6k$ was discussed in the
analysis of \cite{max95}. This may be justified by
demanding that the Kac-Moody currents transform with conformal
weight zero. This is not currently the case, and may be corrected
by adding an ``improvement'' term to the energy momentum tensor. 
This is always possible since the Virasoros are ambiguous up to the addition
of diagonal elements of the algebra and derivatives thereof. The 
general shift for $L_n$ implies that $\De L_0=c_{cl}/24$,
and this boosts the total central charge appearing in the
Virasoro algebra to $c_{tot}=3k/(k+2)+6k\sim 6k$ which agrees
with the asymptotic conformal symmetry of \cite{bh}. Of course, this
modification has not changed the Hilbert space, as we have simply shifted
the lowest eigenvalue of $L_0$ \cite{carlip98}. Thus we see that this shift is
indeed the same as accounting for the zero mode contribution to $L_0$ 
associated with the
classical black hole solution. In this sense this description does not
resolve the microstates, and the asymptotic level density is still described
by a central charge of $O(1)$.

In other words, in the basis
$(\ta,a)$ which we are using, the definition of the characters
under which they form a representation of the modular group
corresponds to (\ref{modchi}), and with reference to the Virasoro
specialisation, it is then the presence of $a\neq 0$
which provides the shift of the central charge to
$c_{tot}=c_q+c_{cl}$. In a basis in which $a$ is also set to zero
one then recovers the Virasoro specialisation of the character
(\ref{vspec}) associated with the combination $L_0-c_{tot}/(24)$.
In other words recovering modular invariance via the prefactor
(\ref{modchi}) is analogous in this basis, to the shift
$L_0\rightarrow (L_0-c/(24))$ in the Virasoro specialisation
which is known to restore modular invariance. In particular, we see that the
Euclidean saddle point is predominantly determined by the ``classical''
zero mode of the black hole gauge field configuration. This analysis 
also indicates
the connection between the choice of $b$, and the shift 
of the lowest eigenvalue of $L_0$ as mentioned above.

\section{Discussion}

To summarise, we have shown that a generic 2+1 cananical decomposition
in the Chern-Simons formulation for Euclidean toroidal geometries admits
a natural modular invariant partition function arising from the affine
symmetry of these solutions. The grand canonical ensemble thaus naturally
contains a modular invariant spectrum of saddle points whose dominance
depends on the position in moduli space, ie. the choice of $\ta$.
Before making some further concluding remarks on the semi-classical limit,
we briefly return to the issue 
of the choice of foliation coordinate
$u$, that was fixed to be the contractible cycle in Section~3 in order to 
focus on the role of the zero mode. If we choose this 
coordinate instead as the non-contractible cycle, 
then the ``spatial'' integral includes no non-contractible
cycles, and it appears that
one can set $g_0=${\bf 1} in (\ref{decomp}). 
However, the global structure of the space
essentially ensures that this zero mode is recovered. We may see this by
noting that there is a class of gauge transformations, leaving
the boundary conditions invariant, which have the following form
\cite{lab}
\be
 \hat{g}_m & = & \exp\left\{\frac{i}{2\rm{Im}\ta}
        (m\overline{\ta}dz-m\ta d\overline{z})
             \right\}, \label{gm}
\ee
and act as $g\rightarrow \hat{g}_m^{-1}g$, $g_0\rightarrow g_0\hat{g}_m$
on the components of the gauge field in (\ref{decomp}). In this expression
$m$ labels the winding around the contractible cycle of the solid torus
which is appropriate for this angular foliation. Now, in general one should
set $m=0$ since otherwise these transformations would be singular
at $r=0$. However, this is not true if, as in our case, there is a 
Wilson line passing through the ``spatial'' slice \cite{lab}, since 
in this case there is no way to apply the transformation at the origin, 
and thus there is no singularity. Consequently, one is able to take such 
transformations into account, and there is no constraint requiring 
$g_0=${\bf 1}, as it can be modified by transformations such as (\ref{gm}).
Indeed we see that if this choice is made initially, then
$g_0\rightarrow g_0\hat{g}_m$ (\ref{gm}) generates precisely
the transform under $S:\ta\rightarrow-1/\ta, a\rightarrow a/\ta$ of the
black hole zero mode (\ref{bhzero}) if we identify $m=a=\overline{a}=1$.

Thus the full partition function is independent of the choice of foliation
as one would expect. Of course, if one only considers particular saddle
points of the partition function then different foliations are quite
convenient. Indeed, we recall once more the intriguing point
that this zero mode structure is encoded at the level of classical 
saddle points by the 3D Smarr formula, as illustrated in the introduction.

As an illustration of the utility of these results we now focus on the
black hole sector, for which its convenient to set $\htt=\ta$, and
consider $|\ta|\rightarrow 0$.
In the Virasoro specialisation, it is a standard argument that
the semi-classical limit of a modular invariant partition function
may be extracted by considering the transformation of
coordinates, $z\rightarrow e^{2\pi i z/\ta}$, $\ta\rightarrow e^{-2\pi i/\ta}$,
appropriate for the $|\ta|\rightarrow 0$ boundary of moduli space
\cite{mod,fs}. This leads to the
saddle points discussed in the introduction. Its interesting that,
although a similar argument based on modular invariance
could be applied to the KM characters discussed here, a 
straightforward extraction of this result follows by studying the 
asymptotic behaviour of the characters for $|\ta|\rightarrow 0$ as follows 
from their explicit representation in terms 
of $\th$--functions \cite{gko,gw,lab}. Recalling (\ref{Zfull}) and 
(\ref{modchi}), we find
\be
  Z_{SL(2,C)}(\ta,a) & \sim & 
    \exp\left(2\pi i \frac{c_{cl}}{24}
                                \left(\frac{1}{\ta}-
       \frac{1}{\overline{\ta}}\right)\right).
\ee
which is the expected semi-classical limit for the
black hole, and we have recalled that $a=1$, and $c_{cl}=6k$.
The calculation may be performed along the lines of that in \cite{bbo},
although we note that the difficulties associated with working
with a negative level $k$, and a non-compact group, can now be
circumvented \cite{maxrev}; details will appear elsewhere. 

One may also observe that to recover thermal $AdS_3$
under a modular transformation, the holonomy $a=1$ should not transform. 
However, this is quite consistent since the thermal $AdS_3$ solution 
needs the holonomy flipped to the other cycle of the torus, and this 
is achieved by ensuring $a$ remains invariant.

Knowledge of the semi-classical saddle points for the partition function,
and the on-shell value of the modular parameter $\ta$,
then allows a straightforward extraction of the entropy, as
$S= (1-\beta\ptl_{\beta})\ln Z$, 
in agreement with the Bekenstein-Hawking value. However, we have
emphasised that the semi-classical contribution to the central charge
which is necessary to describe this density of states is implicit in the
classical background solution, and we have no microscopic picture of the
contributing states. Nonetheless, as we have seen, the WZW picture is still
helpful in unearthing relations between different classical solutions, as
a consequence of an underlying modular structure. We note in passing
that there has been a recent suggestion \cite{ban2} that a more general class
of asymptotic boundary conditions may provide a purely gravitational
description of the black hole microstates. This picture has some connections
with the worldsheet string perspective \cite{gks}, and it would be 
interesting to study the modular structure in this generalised context.
The connection with string theory also requires an extension to
consider the sectors associated with different fermionic 
boundary conditions, for which a
related discussion has recently appeared in \cite{mano}.

\subsection*{Acknowledgments}
We thank M. Ba\~nados for many helpful discussions and comments on the
manuscript, and 
A.R. thanks the Theoretical Physics Group at Imperial College 
for hospitality during
the autumn of 1998 when most of this work was completed.

\bibliographystyle{prsty}

\end{document}